\documentclass[pdflatex,sn-mathphys-num]{sn-jnl}% Math and Physical Sciences Numbered Reference Style
\usepackage{graphicx}%
\usepackage{multirow}%
\usepackage{amsmath,amssymb,amsfonts}%
\usepackage{amsthm}%
\usepackage{mathrsfs}%
\usepackage[title]{appendix}%
\usepackage{xcolor}%
\usepackage{textcomp}%
\usepackage{manyfoot}%
\usepackage{booktabs}%
\usepackage{algorithm}%
\usepackage{algorithmicx}%
\usepackage{algpseudocode}%
\usepackage{listings}%
\usepackage{subcaption}
\usepackage{wrapfig}
\usepackage{caption}
\theoremstyle{thmstyleone}%
\theoremstyle{thmstyletwo}%

\theoremstyle{thmstylethree}%

\begin{document}

\title[Article Title]{Extending the computational reach of Quantum Annealing using Reverse Annealing}


\author*[1]{\fnm{Lucas Joshua} \sur{Menger}}\email{menger@em.uni-frankfurt.de}

\author[1,2]{\fnm{Thomas} \sur{Lippert}}\email{t.lippert@em.uni-frankfurt.de}
\equalcont{These authors contributed equally to this work.}

\author[1]{\fnm{Manpreet Singh} \sur{Jattana}}\email{jattana@em.uni-frankfurt.de}
\equalcont{These authors contributed equally to this work.}

\affil*[1]{\orgdiv{Modular Supercomputing and Quantum Computing}, \orgname{Institute of Computer Science, Goethe University Frankfurt}, \orgaddress{\street{Kettenhofweg}, \city{Frankfurt}, \postcode{60325}, \state{Hessia}, \country{Germany}}}

\affil[2]{\orgdiv{J\"ulich Supercomputing Centre}, \orgname{Forschungszentrum J\"ulich GmbH}, \orgaddress{\street{Wilhelm-Johnen-Straße}, \city{J\"ulich}, \postcode{52428}, \state{North Rhine-Westphalia}, \country{Germany}}}


\abstract{Quantum annealing is a promising heuristic for combinatorial optimization, but on current hardware its performance degrades for larger and more complex problems due to noise and small energy gaps. Reverse annealing has been proposed as a refinement strategy, yet it remains unclear when it provides systematic advantages over standard forward annealing or simply increasing annealing time. We find that combining forward and reverse annealing consistently improves solution quality and efficiency across multiple problem classes. The benefits of reverse annealing increase with problem complexity and are strongest in regimes where forward annealing is increasingly limited. Moreover, reverse annealing yields larger efficiency gains than simply extending forward annealing times. We establish these results through a systematic experimental study on a D-Wave Advantage system, benchmarking reverse annealing across Max-Cut, Number Partitioning, and sparse clustering problems while varying reverse distance, pause duration, and annealing time. We identify a narrow optimal regime for reverse annealing parameters linked to the location of freeze-out points and energy-level crossings in the annealing schedule. These findings demonstrate that reverse annealing is most valuable for large, high-complexity optimization problems and is likely to gain importance as quantum annealing hardware scales toward more realistic applications.}




\maketitle

\section{Introduction}\label{sec1}

Finding optimal solutions to complex combinatorial problems efficiently remains a fundamental challenge in domains such as logistics, finance, and materials design. Yet, many such problems are NP-hard, making exact optimizations on classical systems intractable. While there are many approximate solvers to those problems, they can struggle to find near optimal solutions systematically, especially in cases where heuristics fail \cite{Heu1, Heu2, Heu3, Heu4, Heu5}. This computational limitation has driven interest in alternative paradigms, including quantum optimization. Among these, quantum annealing has emerged as a promising metaheuristic, based on the adiabatic theorem, where quantum fluctuations help the system find a global minimum of an unconstrained combinatorial optimization problem more effectively than classical counterparts \cite{QAdevelop1, Adiabatic2,Adiabatic3,QATheory0,QATheory1,QATheory2,QATheory3,QATheory4}. Unlike gate-based quantum systems, current quantum annealers offer larger qubit counts and lower error rates, positioning them as the most mature quantum hardware for practical optimization tasks \cite{QAdevelop1, QAdevelop2}. However, fundamental questions remain regarding how to best utilize and control these devices for real-world performance gains.\\
\\
Forward annealing (FA), the standard form of quantum annealing, has proven effective in solving quadratic unconstrained binary optimization (QUBO) problems. Another variant of quantum annealing is Reverse Annealing, a technique used as a refinement method that aims to improve on existing solution candidates. While both rely on the same underlying quantum dynamics, they differ in their initialization and evolution. Forward annealing begins from a superposition of all qubits and gradually increases the influence of the problem Hamiltonian until qubits settle in a final state corresponding to the problem. In contrast, reverse annealing starts from a known classical candidate solution and reintroduces quantum effects to refine it. The underlying quantum annealing process can be described by a time-dependent Hamiltonian that interpolates between two components: a mixing Hamiltonian $H_M$, which allows the system to explore different states, and a problem Hamiltonian $H_P$ which encodes the binary quadratic optimization problem as an Ising Hamiltonian. The annealing parameter $s(t) \in [0,1] $, defines the relative weights of the two Hamiltonians over the annealing time:

\begin{gather}
    H(t)=A(s(t)) H_M+B(s(t)) H_P,\\
    H_M=-\frac{1}{2} \sum_i \sigma_i^x,\\
    H_P=-\frac{1}{2} \left( \sum_i h_i \sigma_i^z + \sum_{i>j} J_{i,j} \sigma_i^z \sigma_j^z \right).
\end{gather}

\begin{wrapfigure}{r}{0.5\textwidth}
    \centering
    \vspace{-10pt}
    \includegraphics[width=0.5\textwidth]{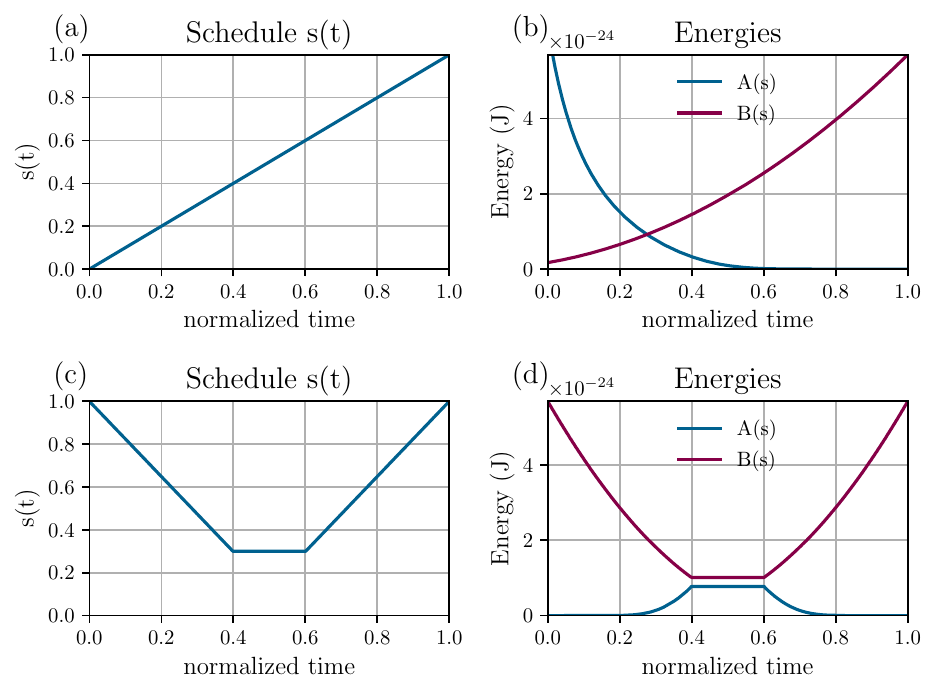}
    \caption{Figures (a) and (b) show the standard version of forward annealing with a linear s(t). Figures (c) and (d) show a reverse annealing schedule with reverse distance 0.3 and pause 0.2. Figures (b) and (d) show the annealing functions for the mixing and the problem Hamiltonian that correspond to the annealing schedules in (a) and (c). Normalized time is the relative time of the annealing process: $t/T$, where $T$ is the total annealing time and $t \in [0,T]$.}
    \label{fig:ann_sched}
    \vspace{-10pt}
\end{wrapfigure}

In forward annealing, the system begins in the easily prepared ground state of $H_M,$ a uniform superposition over all qubit states. As $s(t)$ approaches the value of one, the influence of $H_P$ grows while $H_M$ diminishes. This change in influence is shown by the time dependent annealing functions in b) and d) in Figure \ref{fig:ann_sched}. If the evolution is slow enough to satisfy the adiabatic theorem, the system remains in its instantaneous ground state and eventually reaches the ground state of $H_P$, which corresponds to the solution of the encoded optimization problem.\\
\\
The annealing schedule can be modified by adjusting the shape of the function $s(t)$  and the total annealing time T, where $t \in [0,T]$. $A()$ and $B()$ are fixed annealing schedules controlled through the time dependent function $s(t)$. The interdependency between $s(t)$ and the fixed annealing schedules is depicted in Figure 
\ref{fig:ann_sched}, where a) and b) show a forward annealing example and c) and d) a reverse annealing example. Reverse annealing is a special variation, distinguished not only by a different schedule but also by its initialization from a classical candidate state rather than a superposition state. In this protocol, the system is first prepared in a known classical configuration. The state exploration is then gradually reintroduced by increasing the influence of the mixing Hamiltonian $H_M$. After this exploration phase, the influence of the problem Hamiltonian $H_P$ is ramped up again to guide the system into a potentially improved solution. The reverse annealing schedule is characterized by the depth and duration of the exploration. By using the same time-dependent Hamiltonian framework, reverse annealing enables a local search to potentially escape local minima around the provided initial guess.
The mixing Hamiltonian $H_M$ describes the application of a uniform transverse field on all qubits enabling tunneling between local minima and exploration of the solution space \cite{RAamplitudes, AnnMath}. As the annealing progresses, the influence of the problem Hamiltonian $H_P$ gradually increases, biasing the system toward low-energy configurations corresponding to solutions of the encoded optimization problem. The problem Hamiltonian encodes the objective in terms of qubit biases $h$ and pairwise couplings $J$, both realized by the application of local $z$-fields \cite{RAamplitudes, DwaveQubit}. This Ising formulation can be equivalently represented as a QUBO problem, which we express as the minimization of the quadratic form:
\begin{gather}
    \underset{w \in [0,1]^n}{\text{min}} w^T Q w, \\
    \text{where: } w_i = \frac{\sigma_i^z+1}{2},
\end{gather}
where $Q$ is an $n \times n$ real-valued matrix capturing linear and quadratic coefficients. The QUBO form is particularly useful for the problem formulation and the analysis of its properties.\\
\\
At the end of an anneal, the system is measured in the computational $z$-basis. Before measurement, the anneal schedule reaches a freeze-out point where the system’s dynamics slow down enough that further evolution is effectively frozen \cite{RP6, RP7, RP8, RP9}. Therefore, after the freeze-out point, quantum fluctuations have become negligible, and the qubits have effectively settled into a classical state. As a result, the measurement process does not induce qubit state changes but simply reads out the classical bitstring corresponding to the system's final configuration. This point is problem-dependent and can vary across instances and embeddings. It plays a critical role in reverse annealing, where quantum fluctuations are reintroduced. Reverse annealing is a balancing act between moving beyond the freeze-out point for state exploration while not disrupting promising candidate solutions \cite{M1}.\\
\\
Across different problem classes, and even in constrained settings, forward annealing is able to deliver solutions comparable to those obtained with classical state-of-the-art algorithms. However, as problem sizes grow, hardware imperfections such as device errors increasingly limit performance. These limitations persist even when forward annealing is carefully tuned to the problem and long annealing times are used \cite{annTime, annTime2}. This raises the question of which refinement strategies can enhance FA, particularly in cases where its performance begins to decline. One strategy to mitigate these errors is to slow down the annealing process, giving the system more time to evolve adiabatically. Reverse annealing offers a complementary approach by fine tuning error prone results from a forward annealing run. This dual-anneal strategy enables a two-stage optimization process: a broad global search followed by targeted refinement.\\
\\
Prior work \cite{RR1, RR2, RR3, RR4, RR5, RAamplitudes, RA1, RA2, RA3, RA4} has demonstrated that integrating reverse annealing (RA) into optimization workflows can improve performance in several respects, including higher solution quality for fixed problem sizes, the ability to solve larger instances, or reduced time to reach solutions of comparable quality. Existing studies employ two conceptually distinct approaches to reverse annealing: using RA as a refinement of a classically obtained solution, or using RA to refine solutions generated by forward quantum annealing.
In the first approach, reverse annealing is initialized from a classically computed solution, and multiple studies report that RA can efficiently escape local minima and converge to improved solutions \cite{RR1, RR3, RR4}. In the second approach, which is the focus of this work, forward and reverse annealing are combined into a hybrid quantum workflow. Several studies indicate that this strategy can improve solution quality or enable the solution of problem instances that forward annealing alone cannot reliably solve \cite{RR2, RR5}.\\
\\
Despite these encouraging results, a systematic understanding of when and why reverse annealing provides performance advantages remains lacking. Existing studies typically focus on individual problem classes, specific parameter settings, or isolated performance metrics. While the collective body of work suggests that reverse annealing can be beneficial, methodological differences and limited parameter exploration make it difficult to draw general conclusions or to identify the conditions under which reverse annealing is most effective. Two key gaps therefore remain. First, there is no systematic study that evaluates reverse annealing across multiple problem classes within a unified experimental framework, limiting insight into how problem structure and complexity influence its effectiveness. Second, the role of reverse annealing parameters, such as reverse depth, pause duration, and annealing time—has not been explored through broad parameter scans that explicitly link optimal settings to problem features or annealing dynamics.\\
\\
This work addresses these gaps by conducting a comprehensive experimental study across multiple problem classes chosen to span a range of structural and complexity characteristics, combined with a systematic exploration of reverse annealing parameters. The central question we address is whether the combination of forward and reverse annealing should be regarded as a default strategy for quantum annealing–based optimization, providing consistent improvements in solution quality, efficiency, or achievable problem size across diverse optimization problems.
\\
The remainder of this paper is organized as follows. In Section 1, we describe the experimental setup, including the three benchmark optimization problem classes used. Section 2 outlines the methodology for evaluating reverse annealing as a local refinement strategy. In Section 3, we present the experimental results, followed by an in-depth interpretation and analysis in Section 4. Finally, Section 5 concludes the study and discusses implications for future research and quantum annealing applications.

\section{Benchmarking Problems}\label{sec2}

We evaluate the performance of forward and reverse annealing on three distinct combinatorial optimization problems: Max-Cut, Number Partitioning, and Clustering. This selection was chosen to span a broad spectrum of QUBO characteristics, allowing us to systematically assess the behavior of quantum annealing under varying structural conditions.\\
\\
The chosen problems differ in key properties such as problem size, value density and diversity, number of degenerate ground states, presence of constraints, and the likelihood of encountering good solutions. In addition, energy gaps between both optimal and suboptimal solutions, and between valid and invalid configurations. These factors are known to influence annealing performance and freeze-out behavior.
For each problem class, we build on established QUBO formulations from the literature \cite{MC1, Heu3, QUBOS1, NP1, QUBOS2, CL1, CL2}, making adaptations to improve compatibility with quantum annealing hardware. In the following, we define each problem in detail and outline the QUBO mapping used in our experiments.

\subsection{Maximum Cut}\label{sec:mc}
The Max-Cut problem serves as a baseline in our study due to its simplicity of formulation and its compatibility with quantum annealing hardware. Given an undirected graph $G=(V,E)$, the objective is to partition the set of nodes $V$ into two disjoint subsets such that the number of edges crossing the cut is maximized \cite{MC1, Heu3}.\\
\\
In the QUBO formulation, each node $N_i \in V$ is represented by a binary variable $w_i \in \{0,1\}$ or equivalently a qubit on the QPU. The value of $w_i$ determines the subset to which node $N_i$ belongs. An edge $(N_i,N_j )\in E$ contributes to the cut if $w_i \neq w_j$ and the total cut value can be expressed as a quadratic form over binary variables.
The corresponding QUBO formulation is \cite{QUBOS1}:

\begin{gather}
    H_\text{MC}=-\sum_i \text{deg}(N_i) w_i + 2 \sum_{i>j}((N_i,N_j)\in E)w_i w_j.
\end{gather}

The Max-Cut problem is particularly well-suited for quantum annealing as it requires no auxiliary constraints or variable encodings, every sampled bitstring corresponds to a valid solution. This reduces the impact of bit-flip errors, since they do not invalidate a solution but only slightly lower its quality, thereby increasing the likelihood of finding solutions close to the optimum. All Max-Cut instances used in this study are generated using the Erdős–Rényi model $G(n,p)$, where edges are included with probability $p$, resulting in sparse random graphs \cite{ErdosRenyi}. Sparsity is desirable in quantum annealing as it minimizes the number of required couplers, allowing larger logical problem sizes to be embedded on the hardware.\\
\\
To evaluate solution quality across instances of different sizes and densities, we use a normalized performance metric $q_{\text{MC}}$:

\begin{gather}
    q_{\text{MC}}=\frac{c}{\frac{m}{2}+\frac{n-1}{4}}-1,
\end{gather}

where $m$ is the number of edges, $n$ the number of nodes, and $c$ the number of cuts. This metric expresses the relative improvement over a known upper bound and enables meaningful comparison across different graph realizations \cite{MC1}. Due to the symmetry of the Max-Cut objective, all solutions are two-fold degenerate: flipping all spins yields an equivalent partition. This degeneracy may be beneficial in reverse annealing, as multiple qubit configurations correspond to the ground state, increasing the chances of transitioning into one of these states from a broader set of initial configurations.

\subsection{Number Partitioning}\label{sec:np}

The Number Partitioning problem complements Max-Cut by introducing other structural features: it results in a fully connected QUBO graph that involves bit encodings leading to the requirement of explicit constraint enforcement. In its classical form, the task is to partition a given set of integers into $n$ disjoint subsets such that the absolute difference between the subset sums is minimized. In our experiments, we focus on the case with three subsets, a known NP-hard problem \cite{NP1}.\\
\\
In addition to the objective term that minimizes the absolute difference in subset sums, the QUBO formulation includes a constraint term to enforce valid subset assignments. Subset selection is implemented using a one-hot encoding scheme, where each number is represented by three qubits, and exactly one qubit must be in the "one" state to indicate its assigned subset \cite{QUBOS2}. To enforce this constraint, penalty terms are added to the QUBO that energetically penalize any assignment with multiple or no active qubits for a given number. The QUBO is then a linear combination of both terms where the weight of the constraint decides over the magnitude of the penalty if violated. 
The QUBO representation is given by:

\begin{gather}
    H_p= \sum_{0<j_1<j_2<n+1} \left( \sum_i s_i x_{i,j_1} -\sum_is_i x_{i,j_2} \right)^2, \\
    H_c= \sum_i\left( \sum_j x_{i,j}-1 \right) ^2, \\
    H_\text{NP}= H_p + \lambda H_c.
\end{gather}

Ultimately, we want a weight that distinguishes valid and invalid solutions while it distorts the problem as little as possible.\\
\\
The fully connected nature of the Number Partitioning problem results in a dense QUBO matrix. Due to the use of one-hot encoding for subset assignment, each valid solution has six degenerate counterparts arising from bit permutations that represent the same partitioning. The density of the QUBO and the need for additional encoding qubits significantly limit the maximum implementable problem size. On current quantum annealing hardware, the embedding of fully connected logical graphs is constrained to approximately 150 qubits \cite{H1}.Therefore, Number Partitioning is particularly well-suited for analyzing the robustness of annealing-based refinement techniques, such as reverse annealing, under varying degrees of problem hardness. The hardness can be tuned by manipulating the properties of the input set: sets with widely distributed values induce a broader energy landscape with more local minima and a lower connectivity between near-optimal solutions \cite{EnergyLandscape}. In contrast, sets with narrow or uniform distributions produce a more compressed energy spectrum, increasing the probability of reaching near-optimal solutions but reducing the resolution for refinement.\\
\\
To ensure comparability across problem instances with different total set values, all instances are generated to admit at least one perfect solution in which the subset sums are equal. To assess solution quality independently of scale, we employ a relative metric defined as the inverse, normalized deviation from the ideal bin sum. Specifically, let $S_i$ denote the sum of subset $i$ and $\overline{S}$ their equal value, where $j \leq n$ is the subset index. The quality $q_{\text{NP}}$ is then defined by:

\begin{gather}
    q_{\text{NP}}=1-\frac{1}{k} \sum_j\frac{|S_j-\overline{S}|}{\overline{S}}.
\end{gather}

\subsection{Sparse Clustering}\label{sec:cl}

The clustering problem was selected for two key reasons: it is a common task in unsupervised machine learning, and, under the formulation used here, it enables us to tailor the QUBO representation to the hardware graph \cite{CL2}. While clustering problems typically yield dense QUBO matrices, due to the all-to-all pairwise distance comparisons, the structure can be significantly simplified. This is achieved by employing a spectral clustering-inspired approach, which prunes the underlying graph based on data locality \cite{CL1}. We do this to separate its QUBO structure from Number Partitioning. This leads to a sparse connectivity pattern, reducing the number of non-zero QUBO terms and making the problem more compatible with sparse QPU topologies. Despite structural similarities to Number Partitioning, this variant of clustering emphasizes scalability and hardware-aware formulation, allowing for larger problems.\\
\\
To construct the sparse graph for the QUBO formulation, we employ an iterative graph-building algorithm that enforces a fixed node degree. For each data point, edges are added alternately to either the nearest or the most distant unconnected point, selected from the set of nodes not yet linked to the current node. This process is repeated until every point is connected to 15 other points. The resulting graph ensures uniform connectivity, which can be aligned with the hardware constraints of the QPU’s native topology.\\
\\
Specifically, we test clustering on synthetic datasets generated to form three distinct clusters in two-dimensional space. The clusters vary in size, shape, and distance from one another. By generating the data synthetically, we maintain control over the difficulty and characteristics of the test problems, and we also retain access to ground truth labels indicating the true cluster assignment for each data point.
The clustering task is then formulated as an optimization problem, aimed at maximizing the Euclidean distance between clusters while minimizing the distance within each cluster. This objective translates into the following QUBO formulation:

\begin{gather}
    H_p = \sum_{(i,j) \in P} \sum_{(k,l)\in C} x_{i,k} x_{j,l} (2(k = l) -1) d(i,j), \\
    H_\text{CL} = H_p+ \lambda H_c,
\end{gather}

where $P$ is the combinatorial set of point indices and $C$ the combinatorial set of cluster indices. $H_c$ is the constraint Hamiltonian given in Eq. (9) and $d(i,j)$ is the euclidean distance between point $i$ and $j$. While it would be possible to formulate the objective by only maximizing the distance between clusters, including the minimization of intra-cluster distances proves beneficial for quantum annealing. The reason for this is that the dual-objective approach provides a more informative reward signal for the cluster assignment. Thereby aiding the annealer in satisfying the constraint that each point must belong to exactly one cluster. As a result, the one-hot encoding constraint can be implemented with a lower penalty weight in the QUBO, reducing its disruptive influence on the original optimization objective.\\
\\
Since ground truth labels are available for all clustering instances, solution quality can be quantitatively assessed by comparing intra- and inter-cluster distances with those of the ground truth configuration. Specifically, we define the quality metric $q_{\text{CL}}$ as a normalized measure of clustering effectiveness:

\begin{gather}
    q_{\text{CL}}=\frac{\Delta_{in}/\Delta_{out}}{\Delta_{in}^\star/\Delta_{out}^\star},
\end{gather}

where $\Delta$ measures the sum of euclidean distances between points and the asterisk indicates the ground truth. Note that due to outliers in the cluster data generation, it is possible that a found solution exceeds the ground truth solution in terms of the defined quality metric, leading to $q_{\text{CL}}>1$.

\subsection{Quantitative Characterization of Problem Instances}

To better understand how the chosen optimization problems differ, both within their respective classes and across classes, we characterize each instance using a set of structural and statistical properties \cite{Heu1}. These properties will later be used to explain the performance of forward annealing and to identify the conditions under which reverse annealing offers a measurable advantage.\\
\\
We begin with structural metrics, including the problem size (number of logical qubits), the maximum graph connectivity, and the density of the QUBO matrix. We then examine value-based metrics, such as the distribution of QUBO coefficients and their Shannon entropy \cite{P1}. The Shannon entropy for QUBO coefficients is given by:

\begin{gather}
    \mathbf{H}_1 = - \sum_{i}\sum_{j} p_{i,j} \log_{2} p_{i,j},
\end{gather}

where $p_{i,j}$ is the probability of picking the value of the matrix element $i,j$ out of the set of all QUBO coefficients. The intuition behind entropy as a measure is that it is used in information theory to describe how much information is needed to describe a probabilistic system, the more complex it is the higher the entropy. 
\begin{wrapfigure}{r}{0.5\textwidth}
    \centering
    \vspace{-10pt}
    \includegraphics[width=0.5\textwidth]{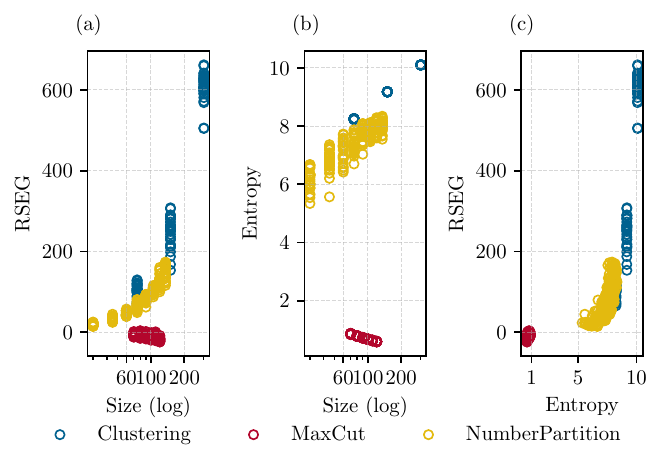}
    \caption{The different approximates of problem complexity and their interdependence. Size is the number of logical qubits of the problem QUBO matrix. The entropy of the QUBO entries is given in Eq. (17). The random state energy gap measures the average gap between the ground truth energy and the energy of a random sampled bitstring.}
    \label{fig:prob_comp}
    \vspace{-10pt}
\end{wrapfigure}
While the QUBO itself is not a probabilistic system the entropy of its variables measures the complexity of their distribution. Finally, we estimate the sampling complexity by comparing the average energy gap between the ground state and configurations obtained from random bitstrings. We call this variable random state energy gap (RSEG). Problems in which random configurations lie closer in energy to the ground state are expected to be inherently easier to solve, whereas larger gaps indicate higher combinatorial complexity.
\\
An important observation one can make in Figure \ref{fig:prob_comp} is the strong correlation between the entropy of the QUBO coefficients and the measured problem hardness. In fact, aside from problem size and class, entropy emerges as the most reliable predictor of hardness across all tested instances.
Moreover, the property distributions summarized in Figure \ref{fig:prob_char} highlight the broad diversity of the benchmark set, spanning a wide range of sizes, connectivities, densities, and value distributions. This diversity ensures that our evaluation covers a representative spectrum of combinatorial optimization scenarios, enabling us to draw more general conclusions about the conditions under which reverse quantum annealing serves as an efficient refinement step following forward annealing.

\section{Reverse Annealing and its Parameters}

Reverse annealing can be implemented through three distinct strategies, each differing in how the initial state is prepared and how the annealing schedule is executed \cite{M1, M2, M3, M4, M5, RP3, M7, RA1, RA2, RA3, RA4}. In the first approach, time-dependent qubit biases are introduced to the Hamiltonian to energetically favor a specific classical configuration as the starting point. These biases are gradually removed during the anneal, while the system transitions from the mixing Hamiltonian toward the problem Hamiltonian, following a modified annealing path similar to forward annealing.\\
\\
The other two strategies, repeated reverse annealing and iterative reverse annealing, do not require modifications to the total Hamiltonian. Instead, they begin by explicitly preparing the system in a specified classical state. The annealing schedule is then executed by first increasing the strength of the mixing Hamiltonian, while lowering the problem Hamiltonian: $\frac{A}{B}\nearrow$. And subsequently reversing the process to return to a classical configuration: $\frac{A}{B}\searrow$, where $A$ and $B$ are the annealing functions in Eq. (1) \cite{RAamplitudes}. 
The key distinction between the two annealing modes lies in how the initial state is handled across multiple annealing runs. In repeated reverse annealing, the system is reinitialized with the same starting configuration for each sample, allowing for statistical sampling of outcomes from a fixed point in solution space. 
\begin{wrapfigure}{r}{0.5\textwidth}
    \centering
    \vspace{-10pt}
    \includegraphics[width=0.5\textwidth]{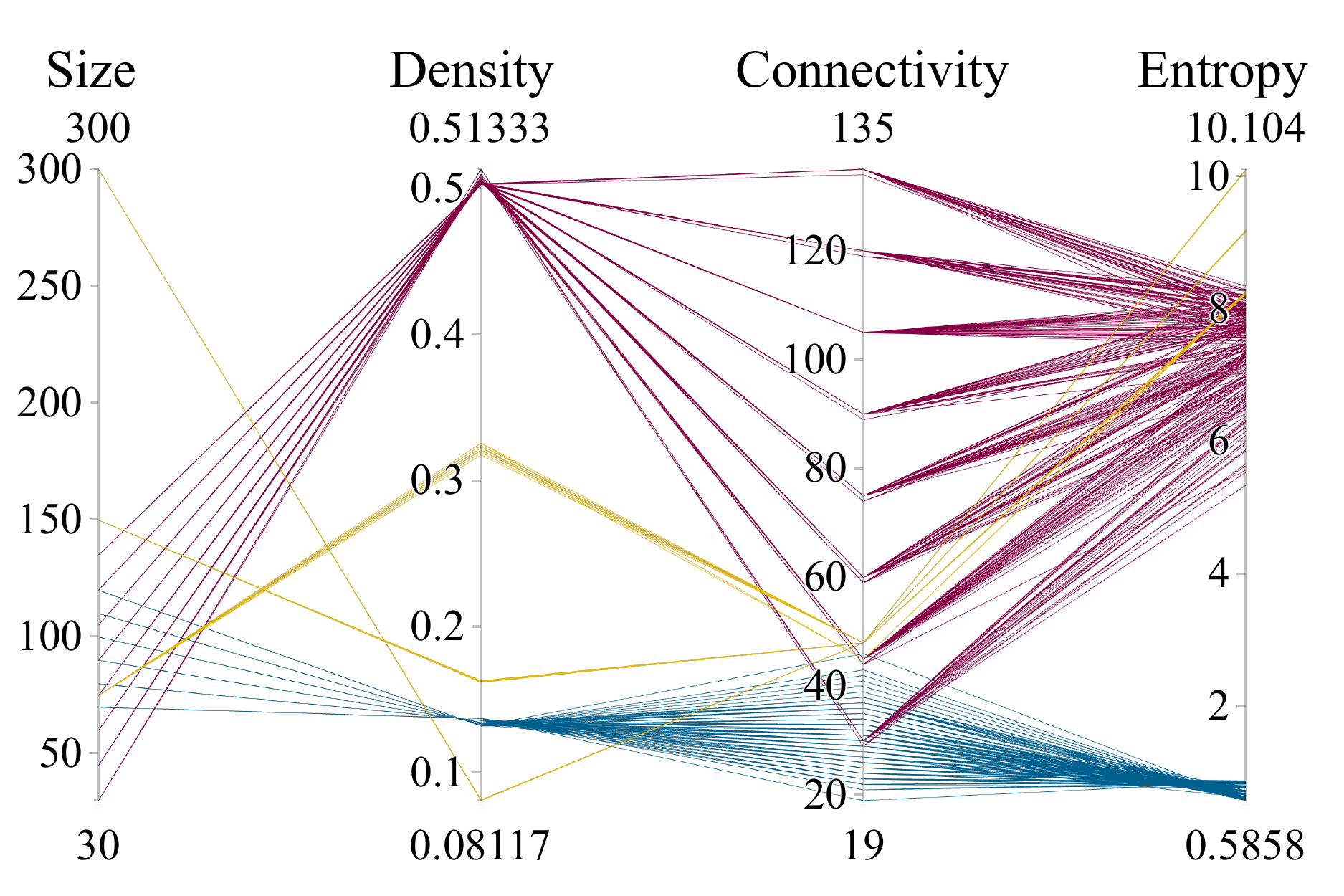}
    \caption{The different characteristics of the QUBO matrix for different problem classes. Size and entropy have the same definition as in Figure \ref{fig:prob_comp}. Density is the relative measure of non-zero values of the QUBO matrix. Connectivity is the number of interconnections when transformed into a graph representation.}
    \label{fig:prob_char}
    \vspace{-10pt}
\end{wrapfigure} 
This approach is well-suited for estimating probabilities of improvement, time-to-solution metrics, and for exploring the robustness of the refinement process under consistent conditions \cite{repeatetExp}. In contrast, iterative reverse annealing uses the final state from one anneal as the starting point for the next. This creates a chain of potentially increasingly refined solutions, following a local optimization path through the solution space. While this may lead to deeper local minima, the changing input state across iterations makes statistical interpretation more difficult. Therefore, we stick to repeated reverse annealing in this study, as our main focus lies in the evaluation of the method and its statistical properties.\\
\\
To understand how reverse annealing can serve as an effective local refinement method, we analyze the eigenenergy spectrum of the annealing Hamiltonian during forward annealing. We consider small Max-Cut instances for which the full Hamiltonian spectrum can be computed exactly. For each point in the annealing schedule (parameter $s$), we calculate the eigenenergies and identify points where different energy levels approach and cross.\\
\\
In Figure \ref{fig:annealing_spec} the background heatmap shows the density of such crossings across 10 small Max-Cut instances, while the line plot illustrates the energy spectrum for a single example. Problems with a well-separated ground state (large energy gaps) exhibit fewer crossings and are less prone to Landau-Zener transitions, but are also harder to improve via reverse annealing. Conversely, systems with smaller gaps and more frequent crossings are generally harder to solve using forward annealing but provide more opportunities for reverse annealing to “hop” from a near-optimal state onto the ground-state path. This usually is true for larger problems \cite{RP1, RP2, RP3, RP4}.As suggested by Fig. 4, the reverse distance controls the number of energy-level crossings and, consequently, the likelihood of transitions to other energy levels. Good reverse distances are therefore expected to lie below 0.5 for the problem classes in this study.\\
\\
With the selected approach of repeated reverse annealing we still have some degrees of freedom that determine the procedure. Aforementioned annealing schedule allows for much variation when it comes to the shape of $s(t)$; we restrict its definition to two parameters: reverse distance and annealing pause. The reverse distance decides on the degree of the reintroduction of the tunneling Hamiltonian. It is measured in terms of $s$, where a smaller $s$ corresponds to a larger influence of the tunneling Hamiltonian. The annealing pause defines the time that the change of the annealing functions $A$ and $B$ is paused. In our case this pause happens in the middle of the annealing process and at the minimum of $s$. We introduce these restrictions to get a symmetric reverse annealing schedule that only depends on parameters deemed to be impactful on the solving success by previous studies \cite{RR5, Pelofske2024}. Both parameters can be summarized as stronger and longer influence of the tunneling Hamiltonian. We expect the reverse distance to be the more influential factor, as it determines whether crossings can occur by closing the energy-level gaps, while the pause primarily increases the duration for which the system is held in this state.\\
\\
\begin{figure*}[ht]
    \centering
    \includegraphics[width=\textwidth]{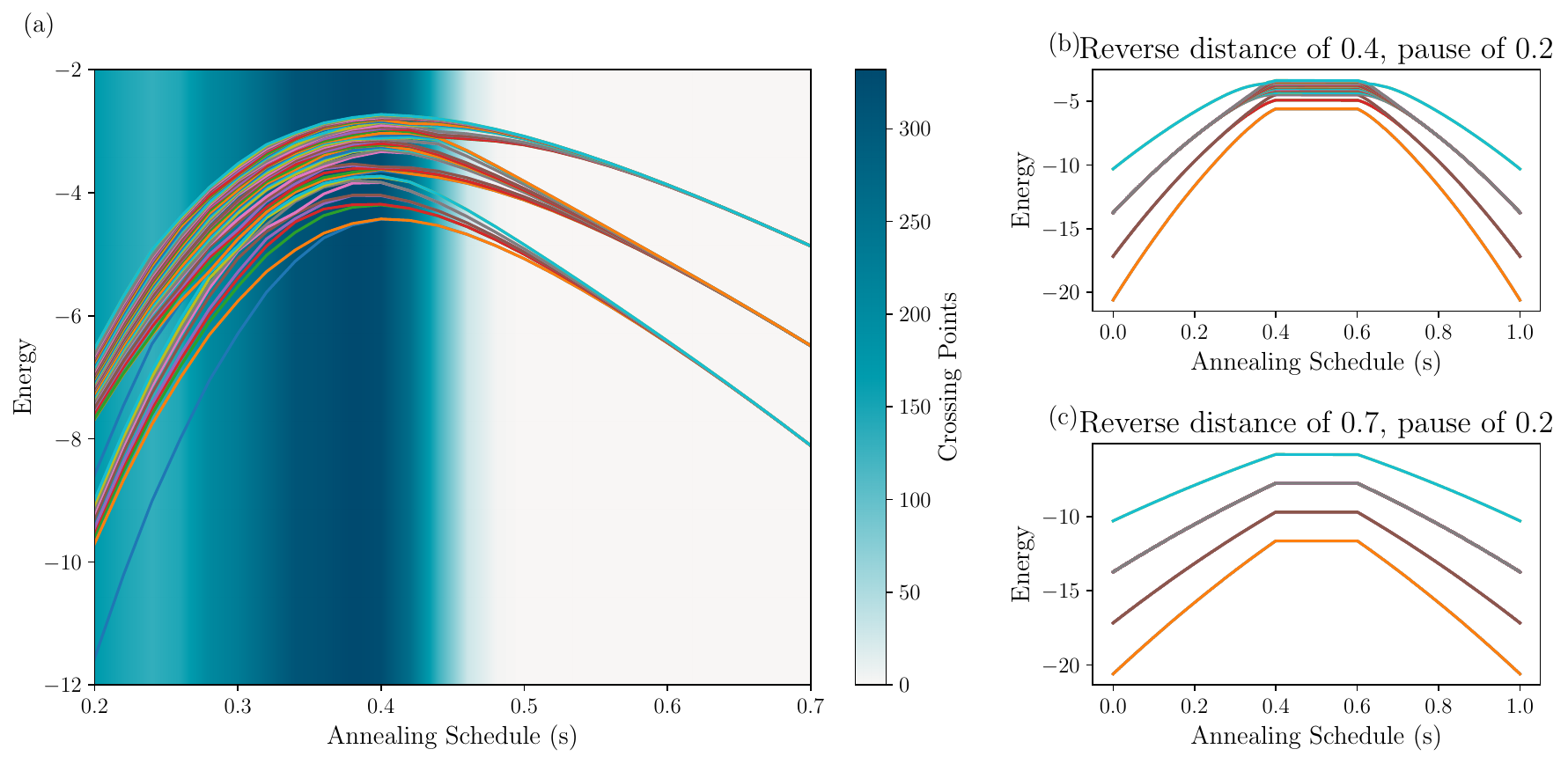}
    \caption{The Figure shows the energy spectrums for forward and reverse annealing. In a) the energy spectra of a forward annealing run for a small Max-Cut instance with 12 qubits is displayed, the background heatmap shows crossing density derived from a set of 10 Max-Cut problems of the same size. The graphs b) and c) show energy spectra for the same problem but for different reverse annealing schedules.}
    \label{fig:annealing_spec}
\end{figure*}
The annealing time is a central element of quantum annealing in general as it is embedded in the adiabatic theorem. Which implies that longer annealing times increase the likelihood of good solutions \cite{annTime, annTime2}. This relationship is not linear as longer annealing times also increase the likelihood of decoherence and other errors. Still, it can be observed that longer annealing times give better solutions in practice \cite{M7}. For two reasons we include annealing time as a parameter; Firstly we want to investigate possible efficiency gains through combinations of forward plus reverse annealing, compared to gains from lengthen the annealing time of forward annealing. Secondly, we want to investigate under which annealing times reverse annealing is more beneficial.\\
\\
During a solving process annealing is repeated multiple times, this process is called sampling, the number of samples is a tunable parameter. With more samples the probability of the occurrence of the optimal solution increases. We fix the number of samples to 1000 in this study, as we assess reverse annealing results from a probabilistic point of view, the number of samples has no influence on the probability of one shot resulting in the optimal solution.\\
\\
QUBO problems often cannot be directly mapped on the QPU as the connectivity between qubits is limited, this makes it necessary to form logical qubits from multiple physical qubits. To ensure that those logical qubits stay coherent, an additional term is needed: the chain strength. It can be seen as an additional constraint term, ensuring that all physical qubits belonging to one logical qubit behave coherently. This is ensured by a strong negative coupling between them, making a correlated behavior energetically more favorable \cite{H1}.

\section{Experimental Results}

To evaluate reverse annealing as a refinement method, we begin by analyzing performance metrics across different conditions for each problem class. Forward annealing serves as the baseline against which all results are compared. We then assess the improvement potential of reverse annealing by examining the relative changes in these metrics, thereby quantifying its effectiveness in enhancing solution quality and efficiency. The compared metrics are the quality metrics described above and time-to-epsilon (TTE) \cite{BM1}:

\begin{gather}
    \mathrm{TTE}_\epsilon=t \, \frac{\ln(1-\alpha)}{\ln\!\left(1 - p_\epsilon\right)},
\end{gather}

where $t$ is the annealing time, $\alpha$ gives the certainty that at least one sufficient solution is sampled, and $p_\epsilon$ is the probability of sampling a solution exceeding the barrier $\epsilon$. We additionally use an adaptation of TTE that we call time-to-improvement (TTI) by using the baseline quality as $\epsilon$-barrier. It is introduced to provide deeper insight into the efficiency of different improvement techniques. This measure enables us to more precisely evaluate which methods yield efficiency gains and whether longer annealing runs are themselves efficient.

\subsection{Forward Annealing}

\begin{figure*}[ht]
    \centering
    \includegraphics[width=\textwidth]{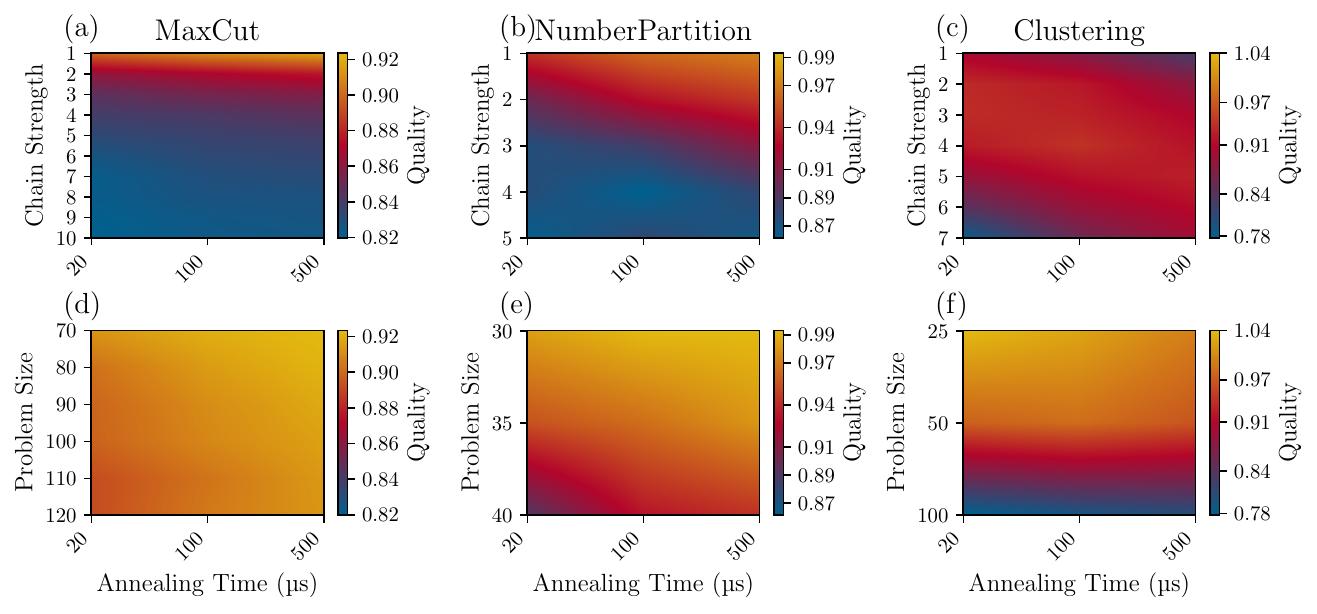}
    \caption{The heatmaps show the average quality metrics defined in Sections \ref{sec:mc}, \ref{sec:np}, \ref{sec:cl} for the three problem classes in dependence of problem size, chain strength and annealing time. For each heatmap, all parameters not shown are averaged over. The displayed quality values correspond to the normalized $q$ metric defined in the previous section.}
    \label{fig:six_fa_qual}
\end{figure*}

The forward annealing experiments serve a second purpose, besides setting the baseline for reverse annealing, to determine the optimal chain strength for forming logical qubits. In the following, we analyze the solution quality for each problem class as a function of problem size, annealing time, and chain strength. To this end, we generated 30 random problem instances per problem class and size. The problem sizes were chosen as follows: for Max-Cut: $ n \in \{ 70, 80, 90, 100, 110, 120 \} $, for Number Partitioning: $ n \in \{ 30, 35, 40 \} $, for Clustering: $ n \in \{ 25, 50, 100 \} $. For each solving process, we performed 1000 annealing runs (samples) on the D-Wave Advantage system, for annealing times ranging from 20 to 500 microseconds. The tests were conducted using the default annealing schedule, with chain strengths varied between 1 and 10.\\
\\
In Figure \ref{fig:six_fa_qual} the results of the baseline experiments are shown. Across the three problem classes, a positive dependence between chain strength and problem size emerges. For Max-Cut, the results strongly favor low chain strengths, largely independent of problem size. For Number Partitioning, this behavior begins to change, with higher chain strengths becoming increasingly favorable for longer annealing times. This trend continues for the clustering problems, where higher chain strengths are generally beneficial and increase with annealing time. A similar but less pronounced shift can also be observed for the Number Partitioning problems. These observations suggest that, for larger instances and longer annealing times, embedding chains become increasingly unstable. Consequently, we face a balancing problem that is highly dependent on both the problem class and annealing parameters, with no universal optimal chain strength. Comparable effects, where extended annealing times introduce additional instability into the solution process, are also evident in the subsequent reverse annealing analysis.\\
\\
Across all problem classes, solution quality typically decreases with increasing problem size and improves with longer annealing times, which can be seen in Figure \ref{fig:six_fa_qual}. The size effect can be decomposed into two factors: the inherent increase in problem complexity with larger instances, which manifests differently across problem classes (see Section 2), and the growing impact of device noise, as the likelihood of qubit errors rises with the number of involved qubits. We utilize forward annealing experiments not only as a baseline for assessing solution quality, but also in combination with the time-to-epsilon metric to derive a quantity we term time-to-improvement. This metric is computed analogously to time-to-epsilon, however, instead of defining epsilon as a fixed target solution quality, the threshold is set to the solution quality achieved by the baseline experiment. All improvement metrics are not limited to reverse annealing but can also be defined for forward annealing experiments. This reflects the idea that, rather than applying a refinement strategy, one may generally increase the annealing time to achieve improved solution quality. Consequently, TTI enables a direct comparison of how longer annealing runs improve upon shorter ones. For forward annealing, TTI uses the next shorter annealing time as the baseline. As a result, it is only defined for the following annealing times:  $ \{100 \mu s, 500\mu s \} $.

\subsection{Reverse Annealing}

To evaluate reverse annealing as an efficient refinement method we conduct similar experiments as for forward annealing, where we take the best chain strength for each problem type and annealing time and test it for different combinations of reverse annealing distance and annealing pause. 
\begin{wrapfigure}[22]{r}{0.5\textwidth}
    \centering
    \vspace{-15pt}
    \includegraphics[width=0.5\textwidth]{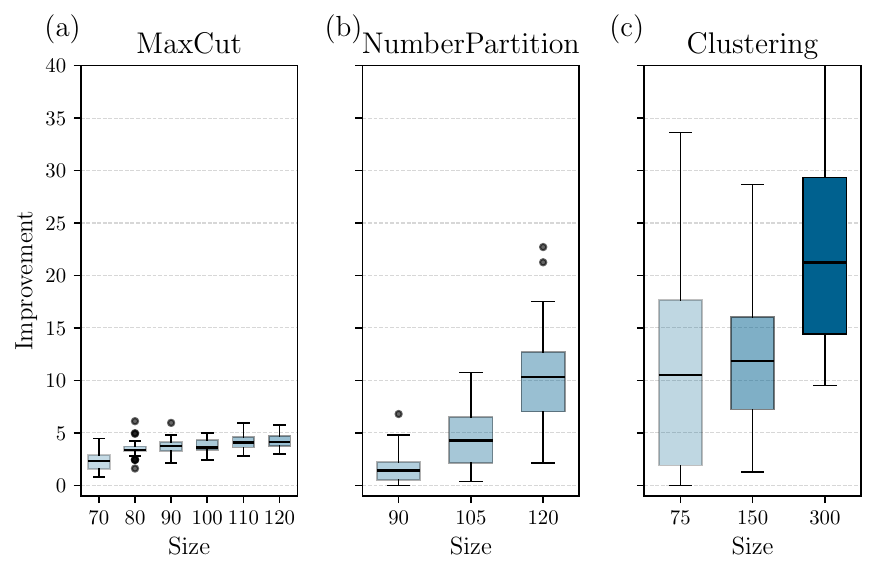}
    \caption{Percentage improvement over the forward annealing baseline that reverse annealing achieves for the three problem classes (a), (b) and (c). The box-plots show the distribution of the improvements in each category, where the box center line is the median, the box outer ends are the upper and lower quartile and the outer end of the line is the 90th percentile. The reverse annealing results are measured from a fixed reverse distance and pause which is selected according to the tuning results given in Table \ref{tab:config}.}
    \label{fig:impr_box}
    \vspace{-15pt}
\end{wrapfigure} 
For each solving process, we performed 1000 repeated reverse annealing runs (samples), for annealing times: $ \{ 20\mu s, 100 \mu s, 500\mu s \} $. The tests were conducted for all combinations of reverse distances and annealing pauses from the two sets: $ \{ 0.2, 0.3, 0.4, 0.5, 0.6 \} \times \{ 0.0, 0.1, 0.2, 0.3, 0.4 \} $. Figure \ref{fig:impr_box} shows the percentage improvement over the baseline of reverse annealing paired with forward annealing across all problem classes and sizes. Moreover, increasing problem complexity appears to have a stronger influence than system size alone, as indicated by the slower growth of improvement for Max-Cut compared to the Number Partitioning and Clustering problems. When compared to the earlier entropy and energy-distance analyses, similar trends are observed. In contrast to those results, the downward slope in Figure \ref{fig:prob_comp} (a) for Max-Cut is not visible here; instead, a mild, positive dependence on size is apparent. This behavior can be attributed to the dominance of resolved qubit errors as the effective problem complexity slightly decreases with increasing size. For the other two problem classes the trends observed in Figure \ref{fig:prob_comp} (a) have the same directions and similar magnitudes as the improvements in \ref{fig:impr_box}, but with differing steepness, likely due to the still-noisy output of reverse annealing. Where Number Partitioning is doubling its improvement for each problem size and Clustering having a small gap between the first two sizes but nearly doubling from the medium size to the largest problems. It is worth noting that for the smallest instances of Max-Cut and number partitioning, there is very little improvement, making the additional annealing time not beneficial from a solution quality perspective.\\
\begin{figure*}[ht]
    \centering
    \includegraphics[width=\textwidth]{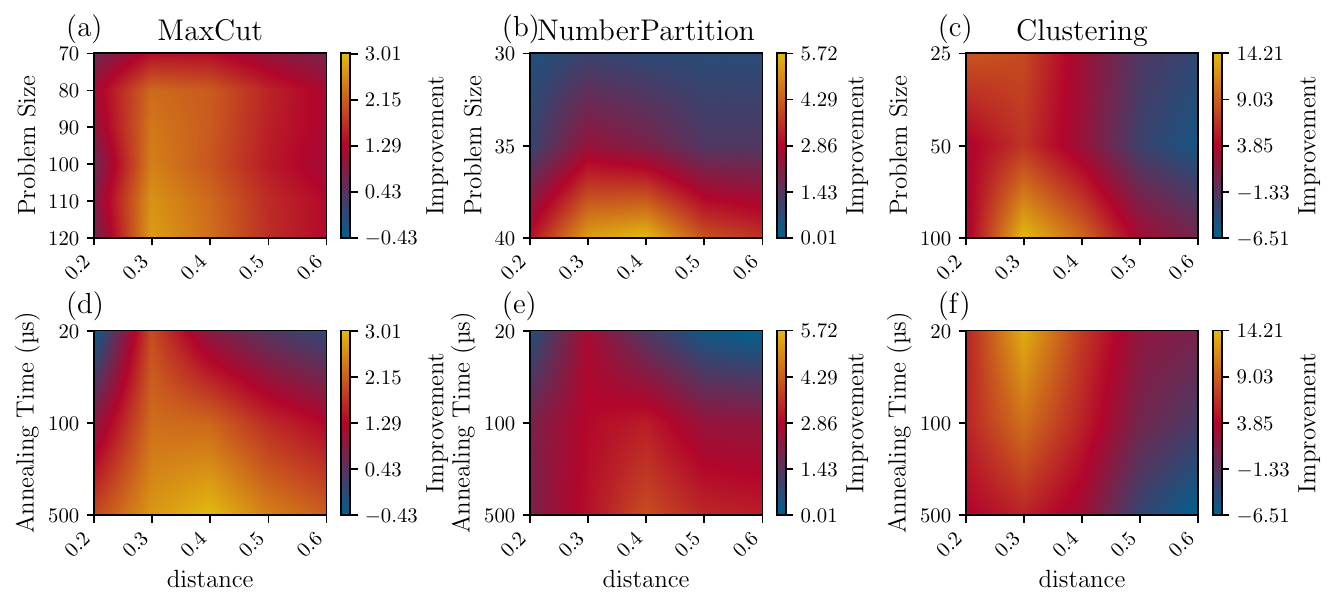}
    \caption{The heatmaps show percentage improvement over the baseline quality in relations to reverse annealing parameters reverse distance, annealing time and problem size. All other parameters not shown (e.g., annealing pause) are averaged over. Negative improvement values indicate cases where reverse annealing failed to produce solutions better than the initial state.}
    \label{fig:six_ra_impr}
\end{figure*}
\\
Figure \ref{fig:six_ra_impr} illustrates the improvement over forward annealing as a function of reverse distance, annealing time, and problem size. The annealing pause is omitted from this figure because it does not contribute significantly to performance gains; its qualitative behavior is similar to that of reverse distance and annealing time, but with improvements an order of magnitude smaller. Among all annealing schedule parameters, the reverse distance emerges as the primary determining factor. Performance improvements are maximized at low reverse distances in the range 0.3 - 0.4. Distances that are too small lead to excessive exploration, resulting in performance degradation or even negative improvement. Conversely, distances that are too large allow only minimal exploration, which may correct isolated bit-flip errors while remaining confined to the same region of the energy landscape, thereby limiting further optimization gains.
\begin{wrapfigure}{r}{0.5\textwidth}
    \centering
    \vspace{-10pt}
    \includegraphics[width=0.5\textwidth]{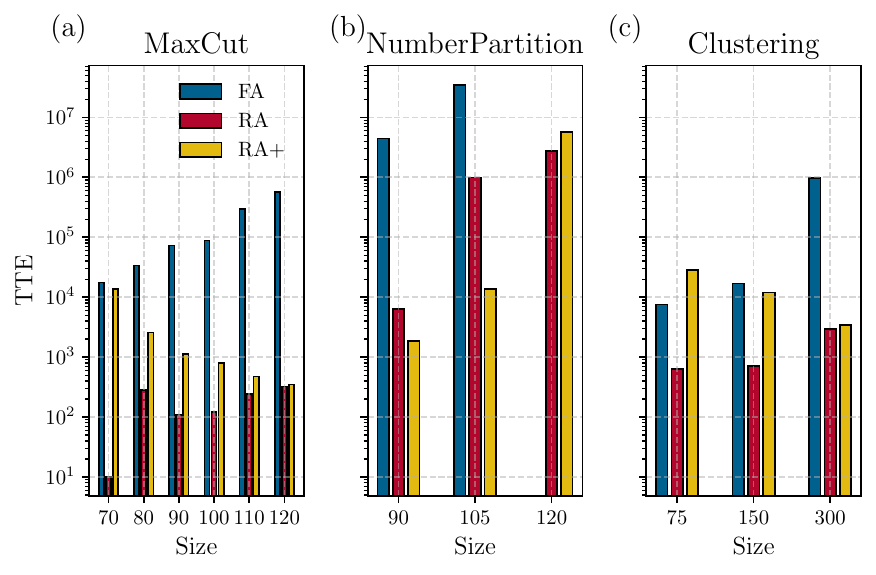}
    \caption{Average time-to-epsilon of forward annealing and reverse annealing for fixed optimal setups. RA+ are the counts of solutions of newly solved problems, that were not solved by forward annealing.}
    \label{fig:tte_bar}
    \vspace{-10pt}
\end{wrapfigure} 
This produces a characteristic horizontal performance ridge centered near a reverse distance of about 0.3. The observed patterns are largely governed by the reverse distance rather than by annealing time or pause duration. Annealing time primarily broadens the range of reverse distances over which improvements are observed, while only mildly enhancing peak performance for the Max-Cut and Number Partitioning problems. For the largest clustering instances, longer annealing times provide no additional benefit.

Table \ref{tab:config} shows the optimal parameter combinations for the different problem classes. These combinations are hereafter referred to as the fixed optimal configuration. This choice is made to illustrate the potential of reverse annealing when it is tuned to the specific problem class. These problems utilize nearly the full capacity of the QPU ($> 4,000$ qubits), where error rates become more pronounced. In this regime, extended annealing times tend to increase the time window for error accumulation rather than enabling a more adiabatic evolution, thereby limiting potential performance gains \cite{errorAccum, errorAccum2}.\\
\begin{table}[t]\footnotesize
\centering
\caption{Summary of the best parameter combinations for reverse annealing combined with forward annealing. $t$ is the annealing time, $s$ is the chain strength, $d$ is the reverse distance and $p$ is the annealing pause. Some of the shown settings deviate for certain problem sizes, for simplicity the best for different sizes is displayed.}
\label{tab:config}
\begin{tabular*}{0.8\textwidth}{@{\extracolsep{\fill}} lcccc}
\toprule
\textbf{Problem Class} & \textbf{$t$} & \textbf{$s$} & \textbf{$d$} & \textbf{$p$} \\
\midrule
MC      &  20     & 1              & 0.3              & 0.0                 \\
MC      &  100      & 1              & 0.3              & 0.0                \\
MC      &  500    & 1              & 0.4              & 0.0                \\
NP      &  20     & 1              &  0.3              & 0.0                 \\
NP      &  100     & 1             & 0.3               & 0.0                 \\
NP      &  500     & 1              & 0.4              & 0.0                \\
CL      &  20     & 3              & 0.3              & 0.0              \\
CL      &  100     & 4              & 0.3              & 0.0                \\
CL      &  500     & 5              & 0.3             & 0.0                 \\
\bottomrule
\end{tabular*}
\end{table}
Quantum annealing is closely tied to considerations of efficiency, with energy-efficient problem solving being one of its central motivations. Consequently, time-to-solution and its variations serve as a key performance metrics, as it directly effects overall energy expenditure. Figure \ref{fig:tte_bar} presents the time-to-epsilon measurements for the different problem classes and sizes, obtained using the optimized annealing schedule parameters and a fixed annealing time of $100 \mu s$. The results show a significantly lower time-to-epsilon for reverse annealing across all problem classes. The improvement is less pronounced for Max-Cut than for the other two classes, primarily because forward annealing already solves many Max-Cut instances effectively, leaving less room for further refinement. In contrast, for the other problem classes the more efficient sampling evident in Max-Cut is combined with the ability to find solutions for previously unsolved instances, resulting in substantially larger efficiency gains. The contribution of discovering new solutions is reflected in the RA+ time-to-epsilon bars.\\

\begin{wrapfigure}{r}{0.5\textwidth}
    \centering
    \vspace{-10pt}
    \includegraphics[width=0.5\textwidth]{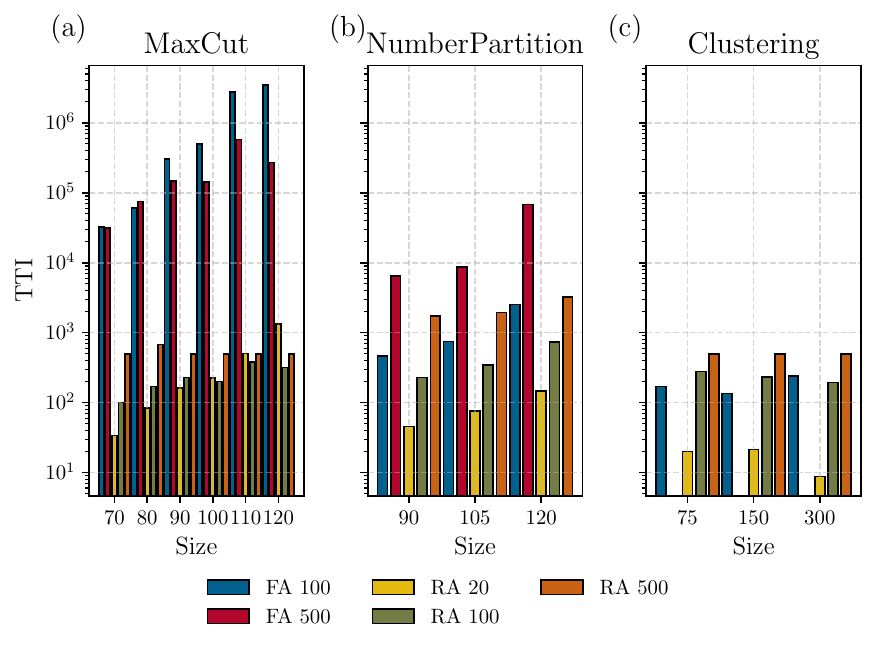}
    \caption{Average time-to-improvement of forward annealing and reverse annealing for fixed optimal setups as defined in this section. FA and RA are the improvement methods and the number is the annealing time in $\mu s$. For FA, the improvement is calculated comparing to the next shorter annealing time. RA is the improvement on FA, both using the same annealing time.}
    \label{fig:tti_bar}
    \vspace{-10pt}
\end{wrapfigure}

\begin{wrapfigure}{r}{0.5\textwidth}
    \centering
    \vspace{-10pt}
    \includegraphics[width=0.5\textwidth]{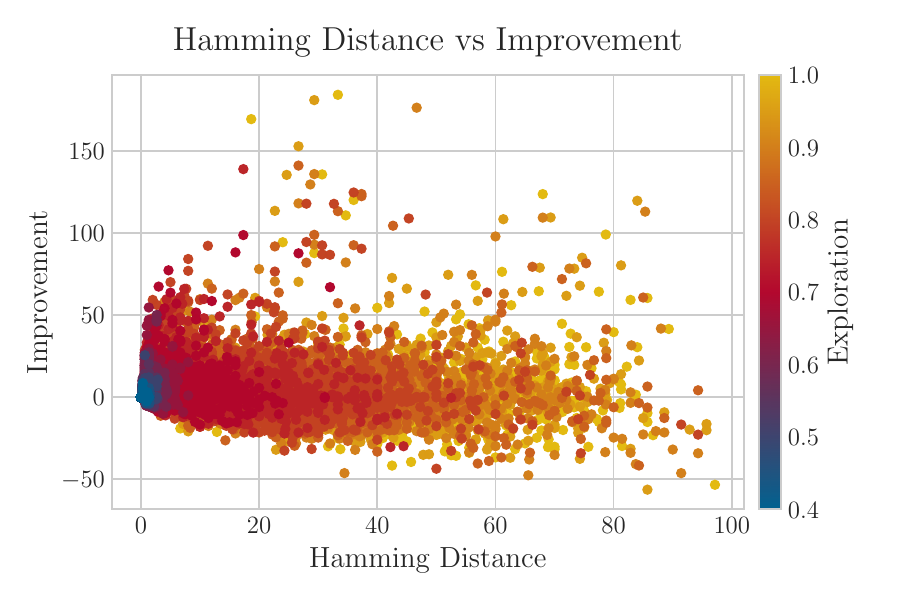}
    \caption{Hamming distance versus improvement with highlighted exploration $e$. exploration is calculated by: $e = 1-d + 0.4 p$, where $e$ is exploration, $d$: reverse distance and $p$: annealing pause.}
    \label{fig:hamming}
    \vspace{-10pt}
\end{wrapfigure}

As the focus of this analysis is the performance benefit provided by reverse annealing, we employ the previously defined time-to-improvement (TTI) metric. Specifically, we compare the efficiency of reverse annealing with that of extended forward annealing in achieving solution improvements over the baseline. This comparison enables an assessment of whether reverse annealing constitutes a more effective refinement strategy than simply increasing the duration of forward annealing runs, both in terms of solution quality and computational efficiency. Figure \ref{fig:tti_bar} displays the resulting differences in TTI between reverse and forward annealing. We observe that solving efficiency increases with longer annealing times and with the use of reverse annealing. This improvement is driven by two complementary effects: obtaining better solutions for problems that were already solvable and discovering solutions for previously unsolved instances. The pattern emerging in Figure \ref{fig:tte_bar} helps to explain the differences observed in the TTI results between the two approaches. The generally more efficient sampling of reverse annealing, combined with its ability to uncover new feasible solutions, enables it to outperform forward annealing in terms of overall solving efficiency.\\
To asses the earlier exploration versus exploitation argument we start by defining an exploration $e$ measure defined by reverse distance $d$ and annealing pause $p$: $e = 1-d + 0.4 p$. We used a parameter scan to identify a weight that optimizes the expressiveness of the exploration measure, yielding a wide range of exploration values per Hamming distance while preserving an approximately linear relationship. This heuristic visualization tool helps to get a clearer understanding of the contributing factors. We then create a scatter plot of improvement compared to hamming distance with colored dots indicating the amount of exploration. The hamming distance is measured between the initial and the final bit string. In Figure \ref{fig:hamming} the results are shown. A clear dependence of exploration and hamming distance is visible, as well as a decline of improvement with higher hamming distances. This shows the effectiveness of reverse annealing as a local refinement method but its limitations on wider, more global searches.

\section{Discussion}
\begin{wrapfigure}[20]{r}{0.5\textwidth}
    \centering
    \vspace{-10pt}
    \includegraphics[width=0.5\textwidth]{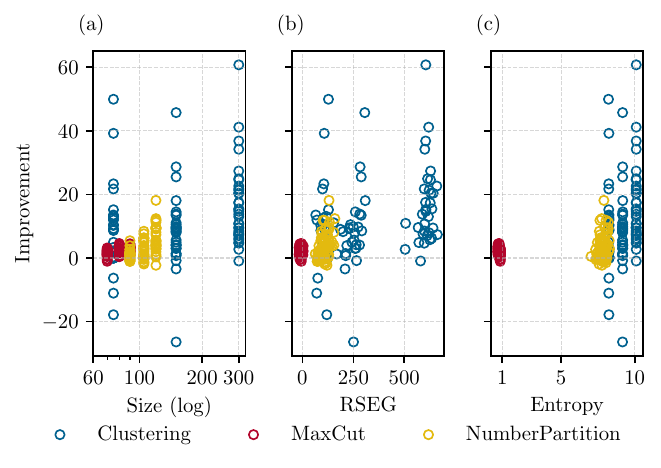}
    \caption{Percentage improvement over the forward annealing baseline, in relation to problem characteristics size (a), random state energy gap (b) and entropy (c), which are the same as in Figure \ref{fig:prob_char}}
    \label{fig:impr_scatter}
    \vspace{-10pt}
\end{wrapfigure}

This section is divided into four parts, each addressing when, why and under which device settings reverse annealing can be effectively used as a refinement technique and when not.\\
\\
Reverse annealing yields the largest efficiency improvements for problems of higher intrinsic complexity. This is evidenced by the strong correlation between efficiency gains and RSEG, as shown in Fig. \ref{fig:impr_scatter}, indicating that reverse annealing is particularly effective in regimes where forward annealing increasingly struggles to reach high-quality solutions. As quantum annealing hardware scales toward larger and more realistic applications, problem complexity is expected to grow, suggesting that the relative advantage of reverse annealing will become more pronounced. To disentangle the contributions of correlated problem characteristics, we performed a linear regression analysis summarized in Table \ref{tab:regression_summary}. While several predictors exhibit substantial interdependence, the analysis indicates that the apparent size dependence of reverse annealing performance is largely mediated by embedding-related effects, most notably average chain length. Similarly, although both entropy and RSEG correlate with performance improvements, RSEG emerges as the more explanatory variable when considered jointly, with entropy serving primarily as a useful proxy for problem hardness rather than a direct driver.
These trends can be understood in terms of the underlying annealing dynamics, in particular the interplay between energy-level crossings and freeze-out. Energy-level crossings occur predominantly before the freeze-out point, enabling transitions that allow the system to move from near-optimal configurations toward lower-energy states. Reverse annealing exploits this regime by reintroducing the mixing Hamiltonian. When freeze-out shifts toward earlier stages, as observed for larger, or more constrained problems, deeper reverse distances become beneficial in order to access these crossings. The mechanisms governing freeze-out behavior, discussed in Refs. \cite{RP5, RP6}, are consistent with the experimental trends observed here. Hardware-related error sources further increase the need for refinement strategies. Qubit errors occur more frequently for problems requiring large numbers of physical qubits, dense and inhomogeneous coupler patterns, or long logical chains induced by minor embedding. Among these factors, chain length plays a particularly prominent role, underscoring the importance of high-quality embeddings and hardware with higher native connectivity for maximizing the benefits of reverse annealing.

Within this context, reverse annealing parameters govern a trade-off between exploration and stability. Longer pause durations reduce the optimal reverse distance by allowing additional time for transitions near closely spaced energy levels, thereby increasing exploration without requiring deeper reversals \cite{RP9}. Conversely, decreasing the reverse distance also enhances exploration by increasing the influence of the transverse field. In both cases, performance is maximized within a narrow regime: excessive exploration, achieved through long pauses combined with deep reversals, disrupts promising candidate solutions and degrades solution quality. A similar trade-off exists between total annealing time and reverse distance, with longer annealing times shifting the optimal reverse distance toward larger values by providing additional time for transitions near energy-level crossings. 
Although combining forward and reverse annealing consistently outperforms forward annealing alone on average, reverse annealing is not universally advantageous. For small or intrinsically simple problem instances, forward annealing already achieves high solution quality with efficient sampling, leaving little room for improvement. In such cases, the additional overhead introduced by reverse annealing may provide negligible benefit and can even degrade performance if the annealing schedule is not carefully tuned. Reverse annealing should therefore be viewed as a targeted refinement strategy, most effective in regimes where forward annealing performance is limited by problem complexity, and due to large embedding graphs, rather than as a default extension of the annealing workflow.

\begin{table*}[!ht]\footnotesize
\centering
\caption{Ordinary Least Squares regression results for improvement variable.}
\label{tab:regression_summary}
\begin{tabular}{lccccc}
\toprule
 & (1) & (2) & (3) & (4) & (5) \\
\midrule
\textbf{Variables} & size & entropy & energy\_distance & chain\_length\_mean & all predictors \\
\midrule
Intercept         & -3.8354*** & 2.3965*** & 3.0840*** & -6.2063*** & 9.5856*** \\
                  & (0.994)    & (0.728)   & (0.528)   & (2.190)    & (2.672)   \\
Size              & 0.0980***  &           &           &            & -0.1619*** \\
                  & (0.007)    &           &           &            & (0.056)   \\
Entropy           &            & 1.2056*** &           &            & -0.7481** \\
                  &            & (0.119)   &           &            & (0.347)   \\
RSEG              &            &           & 0.0395*** &            & 0.1009*** \\
                  &            &           & (0.003)   &            & (0.022)   \\
Chain length mean &            &           &           & 1.1999***  & 0.7495**  \\
                  &            &           &           & (0.181)    & (0.358)   \\
\midrule
Observations      & 364        & 364       & 364       & 364        & 364 \\
R-squared         & 0.327      & 0.222     & 0.402     & 0.108      & 0.418 \\
Adj. R-squared    & 0.325      & 0.220     & 0.401     & 0.106      & 0.412 \\
F-statistic       & 175.6      & 103.3     & 243.6     & 43.85      & 64.51 \\
Prob (F-stat)     & 5.9e-33    & 1.6e-21   & 2.4e-42   & 1.3e-10    & 4.6e-41 \\
\bottomrule
\multicolumn{6}{l}{\footnotesize Robust standard errors in parentheses.} \\
\multicolumn{6}{l}{\footnotesize *** $p<0.01$, ** $p<0.05$, * $p<0.1$.} \\
\end{tabular}
\end{table*}
\newpage
\section{Conclusion}
This work provides a systematic evaluation of reverse annealing as a refinement strategy across multiple problem classes, linking its performance to problem characteristics and annealing dynamics.\\
\\
Across all three problem classes, forward annealing results are consistent with the expectations from theoretical examinations of adiabatic quantum computation \cite{adiabaticPractice1,adiabaticPractice2}. Solution quality decreases with increasing problem size, primarily due to increased probability of qubit errors, while longer annealing times allow slower system evolution and thus improves robustness. Chain strength introduces an additional trade-off: lower values generally preserve the original optimization problem better, provided that chain breaks remain rare.\\
\\
For reverse annealing, the hyperparameter results also align well with theoretical expectations. The reverse distance emerges as the dominant factor for solution quality, as it determines the degree of exploration around the initial state. Optimal performance requires a balance: too much exploration disrupts the refined starting point, while too little prevents meaningful improvement. Pausing introduces additional exploration, leading to interaction effects with reverse distance. Longer annealing times reduce the need for both, as the system naturally explores more of the solution space. Importantly, effective reverse distances coincide with regions of frequent energy spectrum crossings that are located before qubit freeze-out points. This implies that larger problems, with earlier freeze-out points, demand smaller reverse distances.\\
\\
The effectiveness of reverse annealing is strongly tied to intrinsic problem characteristics. As problem size and complexity increase, reverse annealing not only improves solution quality more reliably but also enhances overall efficiency. This trend is particularly evident in its correlation with problem entropy, which, alongside size, emerges as a key predictor of when reverse annealing provides the greatest advantage.
In summary, reverse annealing proves to be an effective refinement strategy across all tested problem classes, consistently improving upon forward annealing baselines. It achieves these improvements efficiently, with competitive time-to-solution and time-to-improvement results. Crucially, the strongest benefits arise for larger and structurally more complex problems, demonstrating that reverse annealing is particularly valuable when problem hardness increases and forward annealing alone becomes less effective.\\
\\
This study is limited in several ways. First, we restricted our comparison to quantum annealing methods and did not benchmark against state-of-the-art classical solvers, which would provide a broader performance context. Second, our efficiency analysis focused exclusively on the annealing process itself, omitting pre- and post-processing steps such as embedding and solution decoding. In practice, the embedding stage proved to be a major bottleneck: its runtime increases rapidly with problem size, and for some instances single embeddings required minutes to compute. This, along with hardware connectivity constraints, limited the maximum problem sizes that could be tested. Finally, our experiments relied on synthetic data, which provides controllable benchmarks but may not fully capture the complexity or structure of real-world problem instances.\\
\\
Future work should extend this study along several directions. First, benchmarking reverse annealing against state-of-the-art classical algorithms would provide a more complete picture of its practical competitiveness. Such comparisons should account for the full solving pipeline, including embedding and post-processing, rather than focusing solely on the annealing step. Second, embedding remains a critical bottleneck for scalability: the runtime of current algorithms grows rapidly with problem size, and in many cases dominated the overall solving process. Developing more efficient embedding heuristics, potentially leveraging machine learning techniques, is therefore a key avenue for progress. Finally, tailoring pre- and post-processing methods specifically for two-step annealing workflows, such as error mitigation for bit flips, could significantly enhance both solution quality and robustness.\\
\\
Taken together, these results establish reverse annealing not merely as a heuristic refinement, but as a systematically effective and physically motivated extension of quantum annealing. By explicitly linking optimal reverse annealing behavior to problem structure, entropy, and annealing dynamics, particularly freeze-out locations and energy-level crossings, this work provides concrete criteria for when and how reverse annealing should be applied. The observed scaling of its benefits with problem size and complexity indicates that reverse annealing becomes increasingly valuable precisely in regimes where forward annealing approaches their practical limits. As quantum annealing hardware continues to scale and tackle more complex optimization tasks, hybrid forward–reverse workflows emerge as a natural and necessary evolution, offering a principled pathway toward improved robustness, efficiency, and solution quality in realistic optimization settings.

\section*{Acknowledgements}
We thank Cedric Gaberle for carefully reading the manuscript and for helpful comments and suggestions.

\section*{Research Funding}
The authors gratefully acknowledge the Jülich Supercomputing Centre (https://www.fz-juelich.de/ias/jsc) for funding this project by providing computing time on the D-Wave Advantage™ System JUPSI through the Jülich UNified Infrastructure for Quantum computing (JUNIQ).

\section*{Ethics Approval}
\textit{Not applicable}

\bibliography{sn-bibliography}% common bib file

\end{document}